\documentclass[onecolumn,preprint,showpacs,superscriptaddress]{revtex4-1}

\usepackage[utf8]{inputenc}
\usepackage{mathtools}
\usepackage{graphicx}
\usepackage{caption}
\usepackage{subcaption}
\usepackage{xcolor}
\usepackage{soul}
\usepackage{float}
\usepackage{amsmath}
\usepackage{amssymb}
\usepackage{amsfonts}

\begin{document}

\title{Klein-Gordon oscillator subject to a Coulomb-type potential in Bonnor-Melvin universe with a cosmological constant}

\author{L. G. Barbosa}
\email{leonardo.barbosa@posgrad.ufsc.br}

\affiliation{Departamento de Física, CFM - Universidade Federal de \\ Santa Catarina; C.P. 476, CEP 88.040-900, Florianópolis, SC, Brazil}

\author{C. C. Barros Jr.}
\email{barros.celso@ufsc.br}

\affiliation{Departamento de Física, CFM - Universidade Federal de \\ Santa Catarina; C.P. 476, CEP 88.040-900, Florianópolis, SC, Brazil} 

\begin{abstract}
In this work we study spin-0 particles described by the Klein-Gordon oscillator formalism in a spacetime which structure is determined by a homogeneous magnetic field and a cosmological constant. For this purpose we take into account a framework based on the Bonnor-Melvin solution with the inclusion of the cosmological constant. We write and solve the Klein-Gordon equation, and then find the energy spectrum by considering the effect of vector and scalar potentials.
\end{abstract}

\maketitle

\section{Introduction}
Nowadays two theories stand out as fundamental pillars of our understanding of physics: the quantum mechanics, that is well known to be an important theory to describe subatomic systems, and the general theory of relativity, that is an important foundation in order to study astronomical systems such as black holes, stars and even the evolution of the Universe. Nevertheless, despite their importance in their typical systems, the relations between these two theories are still not clear enough. 

Currently a substantial amount of investigations have been performed with this purpose. After the study of one electron atoms proposed by Parker \cite{Parker:1980hlc} many works have been made such as the illustrative description of particles in cosmic strings backgrounds, spin-0 fields \cite{Santos:2017eef}, spin-0 fields with the presence of noninertial effects \cite{Santos:2016omw}, rotation effects on scalar fields
\cite{Vitoria:2018its}. Dirac particles near Kerr black holes 
\cite{chandra}, the Klein-Gordon in Schwarzschild spacetime \cite{elizalde_1987}, further instances may be found in   \cite{Sedaghatnia:2019xqb}, that studies Dirac particles in Som-Raychaudhuri spacetime, fermion-antifermion system in a spacetime with topological defect \cite{Guvendi:2022uvz}, Aharonov-Bohm effect for scalar field in a spacetime with a screw dislocation \cite{Vitoria:2018mun}, quasinormal modes of NUT charged black branes
\cite{Cano:2021qzp}, oscillator under rainbow's gravity in wormholes \cite{Guvendi_2023} and WKB decomposition for quantum field theory \cite{Maniccia:2023cgv}.
The Casimir effect \cite{Bezerra:2016brx}, \cite{Santos:2018jba} and bosons in the Hartle-Thorne spacetime near rotating stars \cite{Pinho:2023nfw} are other interesting examples that can be pointed, and all these works provided many interesting results about this question. As it is well known,
an important kind of system which is relevant for many branches of physics is the quantum oscillator. It has been studied in many different formulations in the framework of the general relativity such as in nontrivial topological spacetime with magnetic and quantum flux \cite{Ahmed:2022tca}, \cite{Ahmed:2023blw}, scalar fields in nontrivial topological spacetimes  \cite{Santos:2019izx}, in Gödel-type spacetimes \cite{Yang:2021zxo}, in charged Ellis-Bronnikov wormhole spacetime \cite{Soares:2021uep} or even in the Rindler spacetime \cite{Rouabhia:2023tcl}, presenting a reasonable amount of information. Further investigation in this field brings motivation for this paper.

Strong magnetic fields recently have been observed in magnetars \cite{dunc1995}, 
\cite{Kouveliotou:1998ze} and proposed to be created in high energy heavy ion collisions \cite{kharzeevmag}, \cite{BZDAK2012171}, \cite{voronyuk} and for this reason have become a subject of interest for different kinds of investigations.
In the context of the general relativity some solutions that include magnetic fields are known. Typical examples are the the Bonnor-Melvin solution \cite{WBBonnor_1954}, \cite{MELVIN196465} and a solution for an homogeneous magnetic field which also
includes a cosmological constant ($\Lambda$) \cite{vzofka2019bonnor}. So, it is interesting to study the behavior of quantum particles considering this kind background as it was made in \cite{Konoplya:2007yy}, \cite{Konoplya:2008hj}. In \cite{Santos:2015esa} Dirac particles was studied in the Melvin metric and
in \cite{barbosa2023s}, bosons in the Bonnor-Melvin-$\Lambda$ spacetime. 
In this work we will study quantum oscillators by taking into account spin-0 bosons
described by the Klein-Gordon equation in the Bonnor-Melvin with cosmological constant spacetime and for this purpose we will take into account the metric proposed in \cite{vzofka2019bonnor} with a variation in the definition of the constant of integration in order to provide a clear interpretation for the $\Lambda\rightarrow 0$ limit \cite{barbosa2023s}.

This paper has the following contents: In Sec. II the Bonnor-Melvin with cosmological constant is discussed and in Sec. III the correspondent Klein-Gordon equation is derived, solved and and the energy levels are calculated. In Sec. IV the conclusions of this study are presented.

\section{The Bonnor-Melvin-$\Lambda$ universe} 

 The Bonnor-Melvin with cosmological constant universe is a spacetime which structure is determined by a magnetic field in addition to the cosmological constant.
 This metric is a static solution with cylindrical symmetry of Einstein's equations, generated by a homogeneous magnetic field aligned in the axis direction \cite{vzofka2019bonnor}. For this purpose we will consider the electromagnetic energy-momentum tensor for a potential $A_{\mu}=\left(0,0,0,A_{\varphi}\left(r\right)\right)$ which determines  $F_{\mu\nu}=\nabla_{\nu}A_{\mu}-\nabla_{\mu}A_{\mu}$ that obeys the Maxwell equation $\nabla_{\mu}F^{\mu\nu}=0$.
 So it may be supposed that
 in a cylindrical coordinate system the metric takes the general form
\begin{equation}
    ds^{2}=-e^{A\left(r\right)}dt^{2}+dr^{2}+e^{B\left(r\right)}dz^{2}+e^{C\left(r\right)}d\varphi^{2}
\end{equation}
where $r\geq 0$, $-\infty\leq z\leq \infty$, $0\leq \varphi\leq 2\pi$
and it may be proved that for this configuration 
$A\left(r\right)$ and $B\left(r\right)$ are constants \cite{vzofka2019bonnor}, and then this line element may be written as
\begin{equation}
  ds^{2}=-dt^{2}+dr^{2}+dz^{2}+e^{C\left(r\right)}d\varphi^{2},
\end{equation} 
with 
\begin{equation}
\frac{d^{2}C\left(r\right)}{d r^{2}}+\frac{1}{2}\left(\frac{d C\left(r\right)}{dr}\right)^{2}+4\Lambda=0 \  .
\end{equation} 
A possible solution for this metric is  
\begin{equation} \label{eq:1}
    ds^{2}=-dt^{2}+dr^{2}+dz^{2}+\frac{\sigma^{2}}{2\Lambda}\sin^{2}\left(\sqrt{2\Lambda}r\right)d\varphi^{2}
\end{equation} 
where $\Lambda$ is the cosmological constant and $\sigma$ is a constant of integration.  To determine the electromagnetic potential and, consequently, the magnetic field, we use Maxwell's equation, given by
\begin{equation}
\frac{1}{\sqrt{-g}} \partial_{\nu} \left( \sqrt{-g} \, g^{\alpha\mu} g^{\beta\nu} F_{\alpha\beta} \right) = 0.
\end{equation}
By solving this equation, we obtain the electromagnetic potential as
\begin{equation}
A_{\varphi} = -\sigma \cos\left( \sqrt{2\Lambda} \, r \right),
\end{equation}
which allows us to express the magnetic field as
\begin{equation}
    \begin{split}
H\left(r\right)&=\sigma\sqrt{2\Lambda}\sin\left(\sqrt{2\Lambda}r\right) =H_0\sin\left(\sqrt{2\Lambda}r\right) \\
    \end{split}
\end{equation}
with the intensity of the magnetic field given by $H_0=\sigma\sqrt{2\Lambda}$. We must remark that the integration constant that appears in eq. (\ref{eq:1}) has been defined in a different way if compared with the definition presented in \cite{vzofka2019bonnor}. With this definition he have
\begin{equation}
\lim_{\Lambda\rightarrow0}g_{\varphi\varphi}=\lim_{\Lambda\rightarrow0}\frac{\sigma^{2}}{2\Lambda}\sin^{2}\left(\sqrt{2\Lambda}r\right)=\sigma^{2}r^{2},
\end{equation}
where $\sigma$ determines a conical geometry and is related to a deficit angle $\gamma=2\pi (1-\sigma)$ for a space of constant curvature, with a behavior analogous to a cosmic string spacetime as it is shown in \cite{Santos:2017eef}, \cite{Santos:2016omw}. In this limit, for $\sigma=1$ the Minkowski spacetime is recovered. So, in this work we will study quantum oscillators inside this spacetime.

\section{Klein-Gordon oscillator in the Bonnor-Melvin-$\Lambda$ universe} 
In this section we will formulate the Klein-Gordon oscillator in the background of the
 Bonnor-Melvin with cosmological constant 
 spacetime. For this purpose we will consider bosons described by the Klein-Gordon equation in the metric derived in Sec. II.
This equation may be generalized to curved spaces by \cite{Parker:1980hlc}
\begin{equation} \label{eq:2}
  \frac{1}{\sqrt{-g}}\left(\partial_{\mu}-ieA_{\mu}\right)\left[g^{\mu\nu}\sqrt{-g}\left(\partial_{\nu}-ieA_{\nu}\right)\Psi\right]-m^{2}\Psi=0,
\end{equation}
where $e$ is the charge of the scalar field. We can obtain the Klein-Gordon oscillator, which represents scalar fields in a specific spacetime, following a procedure similar to the minimal coupling performed in the electromagnetism proposed, by Bruce and Mining \cite{bruce1993klein}, and used in other works \cite{Santos:2017eef}, \cite{Santos:2019izx}, \cite{Soares:2021uep} by
changing the momentum operator
\begin{equation} \label{eq:3}
   p^{\mu}p_{\mu}\rightarrow\left(p^{\mu}-im\Omega X^{\mu}\right)\left(p_{\mu}+im\Omega X_{\mu}\right),
\end{equation}
where, \(\Omega\) represents the oscillation frequency, and \(X_{\mu} = (0, r, 0, 0)\). Note that \(X_{\mu}\) is not minimally coupled, which characterizes it as a non-minimal vector interaction. To derive a more general equation, we can also introduce an interaction via a scalar potential \(S(r)\) by modifying the mass term as \(m \rightarrow m + S\). Thus, we consider the most general form of the Klein-Gordon equation with interactions that preserve Lorentz structure, incorporating both vector couplings \(A_{\mu}\), \(X_{\mu}\) and scalar couplings \(S\), so that equation (\ref{eq:2}) becomes:

\begin{equation} \label{eq:4}
\frac{1}{\sqrt{-g}}\left(\partial_{\mu}+im\omega X_{\mu}-ieA_{\mu}\right)\left[\sqrt{-g}g^{\mu\nu}\left(\partial_{\nu}-m\omega X_{\nu}-ieA_{\nu}\right)\Psi\right]-\left(m+S\right)^{2}\Psi=0.
\end{equation}

This equation may be written as
\begin{equation}
    \begin{split}
        &g^{tt}\frac{\partial^{2}\Psi}{\partial t^{2}}+\frac{1}{\sqrt{-g}}\left(\frac{\partial}{\partial r}+m\omega X_{r}\right)\left[\sqrt{-g}g^{rr}\left(\frac{\partial\Psi}{\partial r}-m\omega X_{r}\Psi\right)\right]\\&+g^{zz}\frac{\partial^{2}\Psi}{\partial z^{2}}+g^{\varphi\varphi}\left(\frac{\partial}{\partial\varphi}-ieA_{\varphi}\right)^{2}\Psi-\left(m+S\right)^{2}\Psi=0,
    \end{split}
\end{equation}
and taking into account the symmetry of the problem, we can consider an Ansatz of the form
\begin{equation}\label{fact}
\Psi\left(t,r,\varphi,z\right)=e^{-i\varepsilon t}e^{i\ell\varphi}e^{ip_{z}z}R\left(r\right)
\end{equation}
where $\varepsilon$, $\ell=0,\pm 1, \pm 2, \cdots$ and $p_z$, which is an arbitrary constant, are respectively the quantum numbers associated with energy, angular momentum around $\varphi$ and momentum in the $z$ direction. Replacing the metric coefficients and using eq. (\ref{fact}) we get the radial equation
\begin{equation}\label{eqrad1}
    \begin{split}
      &\frac{1}{\sin\left(\sqrt{2\Lambda}r\right)}\left(\frac{d}{d r}+m\omega r\right)\left[\sin\left(\sqrt{2\Lambda}r\right)\left(\frac{d R\left(r\right)}{d r}-m\omega rR\left(r\right)\right)\right]\\&\qquad+\left[\varepsilon^{2}-p_{z}^{2}-\frac{2\Lambda}{\sigma^{2}\sin^{2}\left(\sqrt{2\Lambda}r\right)}\left(\ell-eA_{\varphi}\right)^{2}-\left(m+S\right)^{2}\right]R\left(r\right)=0
    \end{split}
\end{equation}
which is given in  terms of the quantum numbers and the potentials. 

Observing that the estimated value of the 
cosmological constant is very small and that we are studying quantum particles localized near the symmetry axis, consequently looking for solutions for small values of $r$, 
we can consider an expansion up to $\mathcal{O}\left(2\Lambda r^{2}\right)$ in the components of the radial equation obtaining a new equation that will provide solutions with a high degree of precision. Even if one considers the solar radius, this factor will be of the order of $10^{-34}$ for the usual value of $\Lambda$ \cite{Workman:2022ynf} and in this approximation, eq. (\ref{eqrad1}) becomes

\begin{equation}\label{eqrad2}
   \begin{split}
       &\frac{1}{r}\left(\frac{d}{d r}+m\omega r\right)\left(r\frac{d R\left(r\right)}{d r}-m\omega r^{2}R\left(r\right)\right)\\&+\left(\xi^{2}-\frac{\ell^{2}}{\sigma^{2}r^{2}}+\frac{2\ell e}{\sigma^{2}r^{2}}A_{\varphi}-\frac{e^{2}}{\sigma^{2}r^{2}}A_{\varphi}^{2}-2mS-S^{2}\right)R\left(r\right)=0.
   \end{split}
\end{equation}

An illustrative case, aiming to obtain specific solutions for the equation, is to consider a potential of the Coulomb type, given by $S(r)=\eta / r$, where $\eta$ represents the coupling parameter. In this scenario, the radial equation takes the form

\begin{equation} \label{eq:8}
\begin{split}
   &\frac{d^{2}R\left(r\right)}{d r^{2}}+\frac{1}{r}\frac{d R\left(r\right)}{d r}\\&+\left[\xi^{2}+2\Lambda e\left(\frac{\ell}{\sigma}+e\right)-2m\omega-\frac{2m\eta}{r}-\frac{\left(\ell/\sigma+e\right)^{2}+\eta^{2}}{r^{2}}-\left(e^{2}\Lambda^{2}-m^{2}\omega^{2}\right)r^{2}\right]R\left(r\right)=0
\end{split}
\end{equation}
where we define $\xi^{2}=\varepsilon^{2}-p_{z}^{2}-m^{2}$. Supposing a solution given by $R\left(r\right)=r^{-\frac{1}{2}}\Phi\left(r\right)$  we can write the equation in the form of
an effective Schrödinger one, for an effective potential $ V_{\text{eff}}$,
\begin{equation}\label{EqEff}
 \frac{d^{2}\Phi\left(r\right)}{dr^{2}}+\left(\xi^{2}-V_{\text{eff}}\right)\Phi\left(r\right)=0 \ ,
\end{equation}
where
\begin{equation}
 V_{\text{eff}}=-2\Lambda e\left(\frac{\ell}{\sigma}+e\right)+2m\omega+\frac{2m\eta}{r}+\frac{\left(\ell/\sigma+e\right)^{2}+\eta^{2}-1/4}{r^{2}}+\left(e^{2}\Lambda^{2}-m^{2}\omega^{2}\right)r^{2}.
\end{equation}

Taking the following transformation to the radial effective Schrodinger equation (\ref{EqEff})
\begin{equation}\label{Phi}
 \Phi\left(r\right)=r^{\frac{1}{2}}r^{\sqrt{\left(\frac{\ell}{\sigma}+e\right)^{2}+\eta^{2}}}e^{-\frac{1}{2}\kappa r^{2}}H\left(r\right)
\end{equation}
where $\kappa=\left(e^{2}\Lambda^{2}-m^{2}\omega^{2}\right)^{\frac{1}{2}}$, then the change of variable $x=\sqrt{\kappa} r$ we obtain the Heun's bicofluent equation \cite{ronveaux1995heun}
\begin{equation}
    \frac{d^{2}H\left(x\right)}{dx^{2}}+\left(\frac{1+\alpha}{x}-\beta-2x\right)\frac{dH\left(x\right)}{dx}+\left\{ \left(\gamma-\alpha-2\right)-\frac{1}{2}\left[\delta+\left(1+\alpha\right)\beta\right]\frac{1}{x}\right\} H\left(x\right)=0
\end{equation}
where 
\begin{equation}\label{parameters}
    \alpha=2\sqrt{\left(\frac{\ell}{\sigma}+e\right)^{2}+\eta^{2}},\quad\beta=0,\quad\gamma=\frac{\xi^{2}}{\kappa}+\frac{2\Lambda e}{\kappa}\left(\frac{\ell}{\sigma}+e\right)-\frac{2m\omega}{\kappa},\quad\delta=\frac{4m\eta}{\sqrt{\kappa}} \  .
\end{equation}

The general solution to this equation can be expressed using biconfluent Heun functions \cite{ronveaux1995heun,vieira2015quantum,Cunha:2016uch}
\begin{equation}\label{Heun}
H\left(x\right)=C_{1}\text{HeunB}\left(\alpha,\beta,\gamma,\delta,x\right)+C_{2}x^{-\alpha}\text{HeunB}\left(-\alpha,\beta,\gamma,\delta;x\right),
\end{equation}
where \( C_{1} \) and \( C_{2} \) are constants, and to obtain a well-defined solution at the origin we set \( C_2 = 0 \). Thus, considering (\ref{Phi}) and (\ref{Heun}), the radial equation can be expressed in terms of the Heun bifluent function
\begin{equation}\label{FRadial}
    R\left(r\right)=r^{\frac{1}{2}\alpha}e^{-\frac{1}{2}\kappa r^{2}}\text{HeunB}\left(\alpha,\beta,\gamma,\delta;\sqrt{\kappa}r\right).
\end{equation}

When \(\alpha\) is not a negative integer, the biconfluent Heun functions are given by
\begin{equation}
    \text{HeunB}\left(\alpha,\beta,\gamma,\delta;x\right)=\sum_{j=0}^{\infty}\frac{A_{j}}{\left(1+\alpha\right)_{j}}\frac{x^{j}}{j!}=\sum_{j=0}^{\infty}\frac{\Gamma\left(1+\alpha\right)A_{j}}{\Gamma\left(j+1+\alpha\right)}\frac{x^{j}}{j!},
\end{equation}
where the coefficients \( A_{j} \) satisfy the recurrence relation \( \left(j\geq0\right) \)
\begin{equation}
    A_{j+2}=\left[\left(j+1\right)\beta+\frac{1}{2}\left[\delta+\left(1+\alpha\right)\beta\right]\right]A_{j+1}-\left(j+1\right)\left(j+1+\alpha\right)\left(\gamma-\alpha-2-2j\right)A_{j}. 
\end{equation}
Analyzing this recurrence relation, we observe that the function \( \text{HeunB}\left(\alpha,\beta,\gamma,\delta;x\right) \) becomes a polynomial of degree \( n \) if the following conditions are satisfied
\begin{equation}\label{HeunCondI}
    \gamma-\alpha-2=2n,\quad n=0,1,2,\cdots,
\end{equation}
\begin{equation}\label{HeunCondII}
    A_{n+1}=0,
\end{equation}
where \( A_{n+1} \) has \( n+1 \) real roots when \( 1+\alpha>0 \) and \( \beta\in\mathbb{R} \). This condition can be represented by a three-dimensional diagonal determinant of order \( n+1 \), as demonstrated in \cite{ronveaux1995heun} and may be used to determine the energy spectrum of the system as far as the energy appears in the $\gamma$ factor defined in eq. (\ref{parameters}) .

So, considering the condition (\ref{HeunCondI}), we obtain a discretization condition for the energy levels \cite{Vieira:2016ubt, Vieira:2015dwa, Vieira:2014waa} and carrying out some manipulations we can write
\begin{equation}\label{en}
	\varepsilon_{\pm}=\pm\sqrt{\zeta^{2}+2\kappa\left(n+\sqrt{\left(\frac{\ell}{\sigma}+e\right)^{2}+\eta^{2}}+1\right)-2\Lambda e\left(\frac{\ell}{\sigma}+e\right)}, 
\end{equation}
where $\zeta^{2}=m^{2}+p_{z}^{2}+2m\omega$ and taking into account the relationship between the magnetic field and the cosmological constant $\sigma=H_{0}/\sqrt{2\Lambda}$, and considering the natural system of units in order to express the energy spectrum in GeV, we can write
\begin{equation}\label{en2}
  \varepsilon_{\pm}=\pm\sqrt{\zeta^{2}+2\kappa\left(n+\sqrt{\left(\frac{\sqrt{2\Lambda}\ell}{\sqrt{G}H_{0}}+e\right)^{2}+\eta^{2}}+1\right)-2\Lambda e\left(\frac{\sqrt{2\Lambda}\ell}{\sqrt{G}H_{0}}+e\right)}
\end{equation}
which, as we can see, is influenced by the oscillation frequency $\omega$, by the scalar potential parameter $\eta$, by the intensity of the magnetic field $H_0$ and by the cosmological constant $\Lambda$. We should note that for \( m^{2}\omega^{2} > e^{2}\Lambda^{2} \), the energy spectrum takes on complex values, indicating potential instabilities in the system. More energy levels can be found by analyzing the truncation of the power series for each value of $n$, in the same way as was done in \cite{Leite:2015rjk}, \cite{Leite:2019hgn}.

\begin{figure}[H]
     \centering
     \begin{subfigure}[b]{0.3\textwidth}
         \centering
         \includegraphics[width=\textwidth]{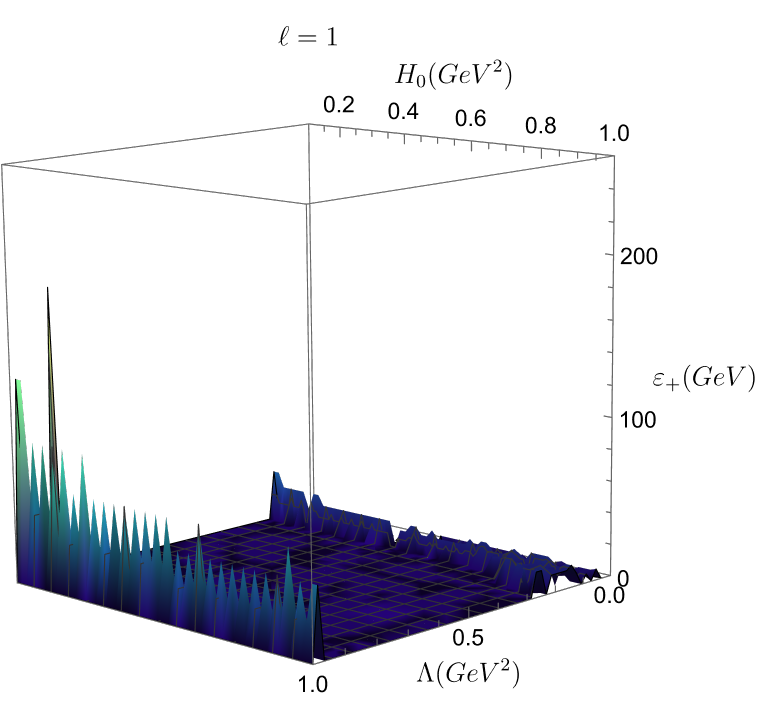}
     \end{subfigure}
     \begin{subfigure}[b]{0.3\textwidth}
         \centering
         \includegraphics[width=\textwidth]{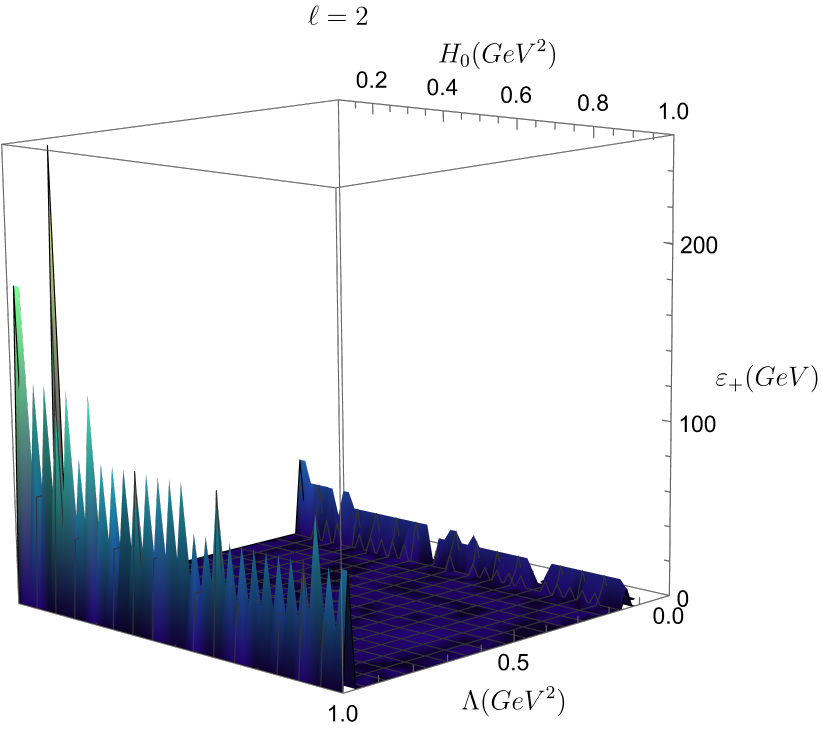}
     \end{subfigure}
    \begin{subfigure}[b]{0.3\textwidth}
         \centering
         \includegraphics[width=\textwidth]{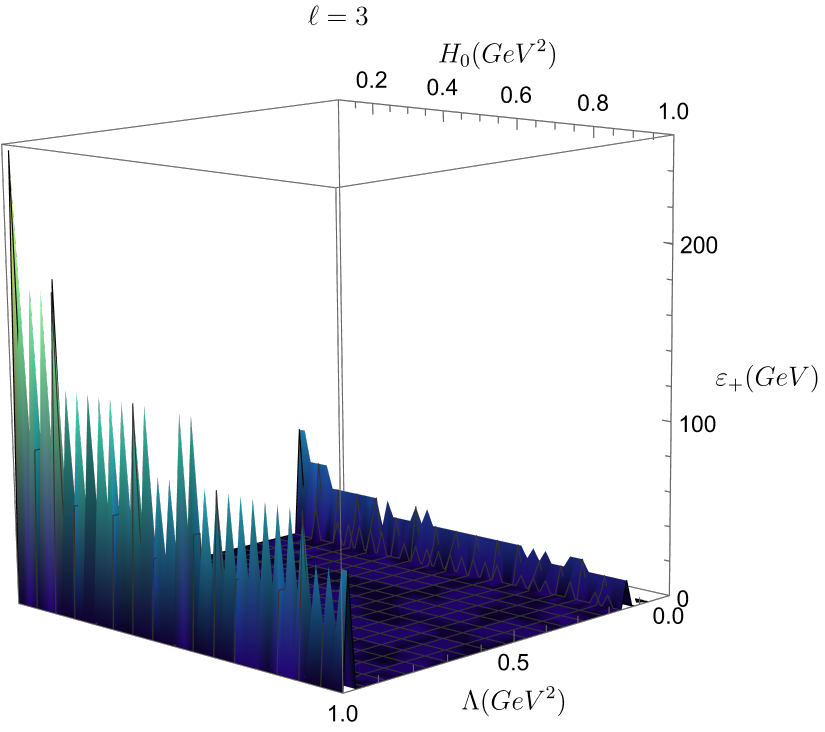}
     \end{subfigure}
        \caption{Graph of the energy spectrum in GeV with respect to the magnetic field intensity $H_{0}$ and the cosmological constant $\Lambda$, for $\ell=1,2,3$, $m=p_z= 0.139$ GeV, $G=6.70 \times 10^{-39}\text{GeV}^{2}$, $e=0.3$, $\eta=0.3$, $\omega=7.25\times10^{-9}$ and $n=1000$.}
        \label{f1}
\end{figure}

\begin{figure}[H]
     \centering
     \begin{subfigure}[b]{0.3\textwidth}
         \centering
         \includegraphics[width=\textwidth]{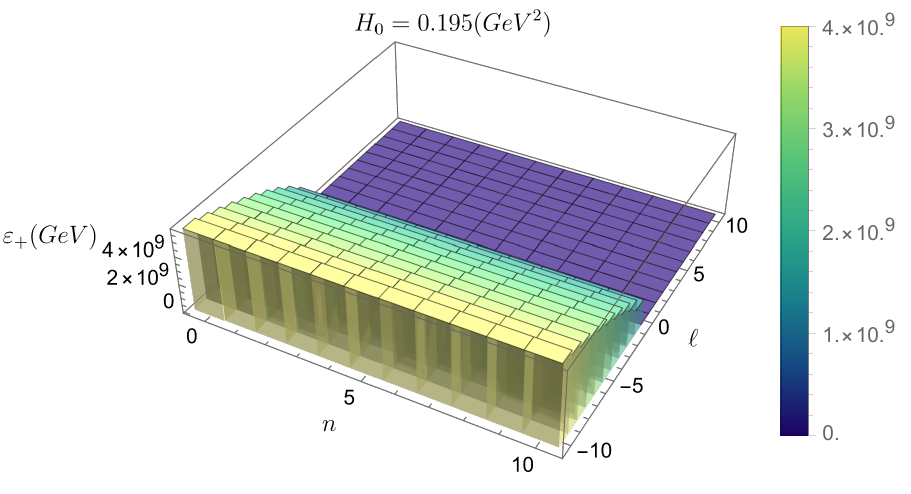}
     \end{subfigure}
     \begin{subfigure}[b]{0.3\textwidth}
         \centering
         \includegraphics[width=\textwidth]{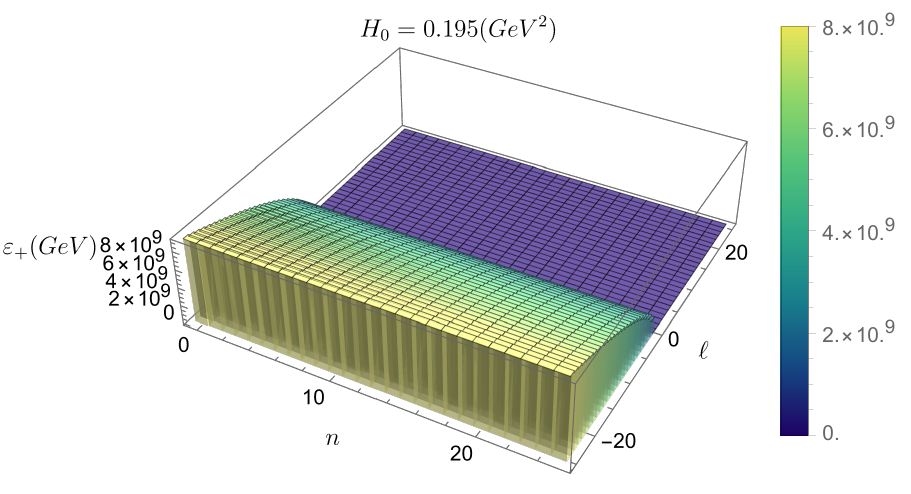}
     \end{subfigure}
    \begin{subfigure}[b]{0.3\textwidth}
         \centering
         \includegraphics[width=\textwidth]{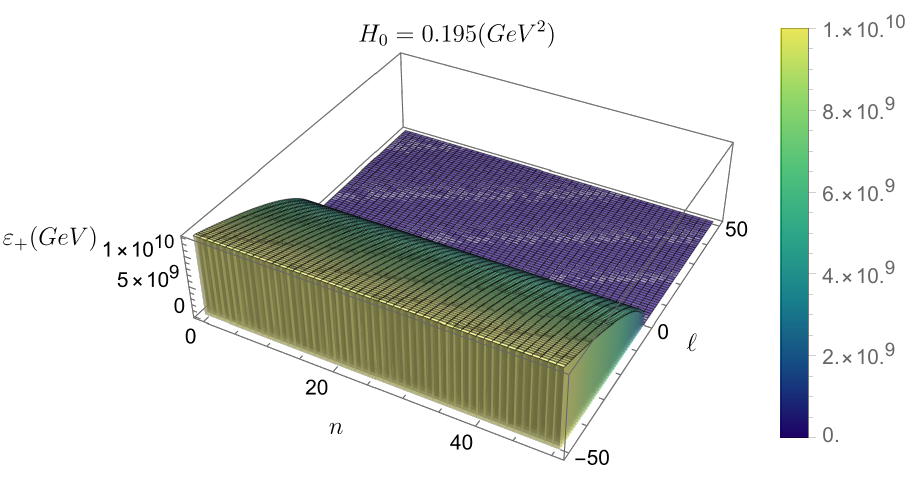}
     \end{subfigure}
        \caption{Plot of the energy spectrum in GeV with respect to the quantum numbers $n$ and $\ell$, with $H_{0}=0.195 \text{ GeV}^{2}$ and the cosmological constant $\Lambda=0.1 \text{ GeV}^{2}$, for $m=p_z= 0.139$ GeV, $e=0.3$, $\eta=0.3$, $\omega=7.25\times10^{-9}$.}
        \label{f2}
\end{figure}

Some numerical results for the energy levels are presented in Fig. \ref{f1} and Fig. \ref{f2} and show their general behavior with respect of the considered parameters.  
In Fig. \ref{f1}, the dependence on $H_0$ and $\Lambda$ is investigated for $n=1000$ and $\ell=1$, 2 and 3 taking the pion mass $m$ as example with $p_z=m$. 
In Fig. \ref{f2}, the dependence with $n$ and $\ell$ is studied for fixed values of $\Lambda$ and $H_0$ for the same set of parameters.

\begin{figure}[H]
     \centering
     \begin{subfigure}[b]{0.3\textwidth}
         \centering
         \includegraphics[width=\textwidth]{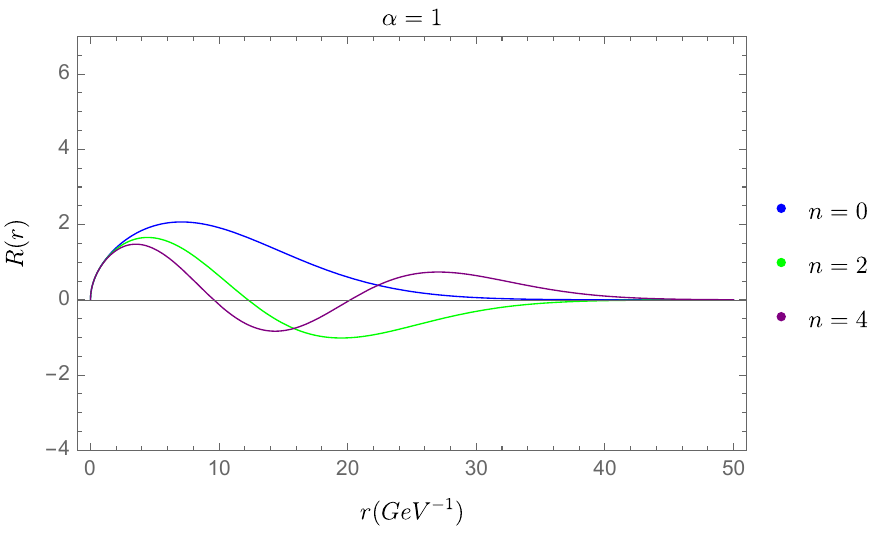}
     \end{subfigure}
     \begin{subfigure}[b]{0.3\textwidth}
         \centering
         \includegraphics[width=\textwidth]{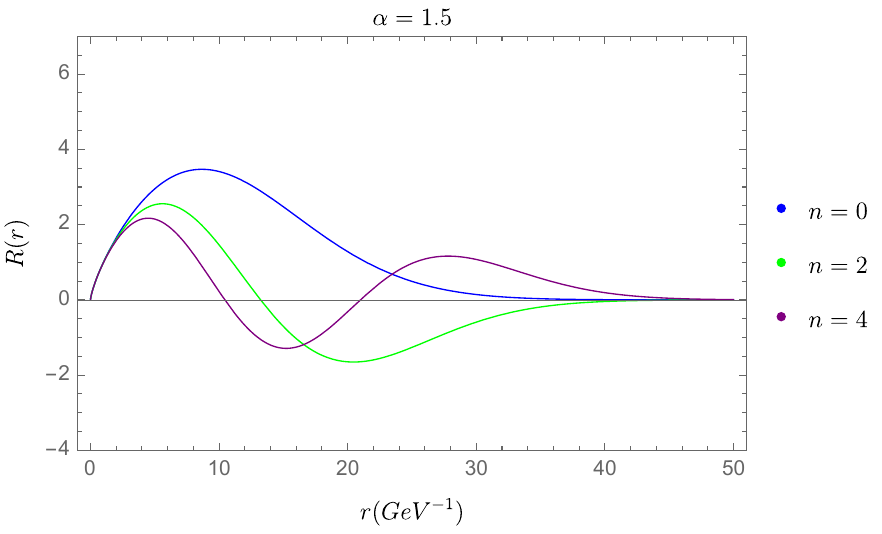}
     \end{subfigure}
    \begin{subfigure}[b]{0.3\textwidth}
         \centering
         \includegraphics[width=\textwidth]{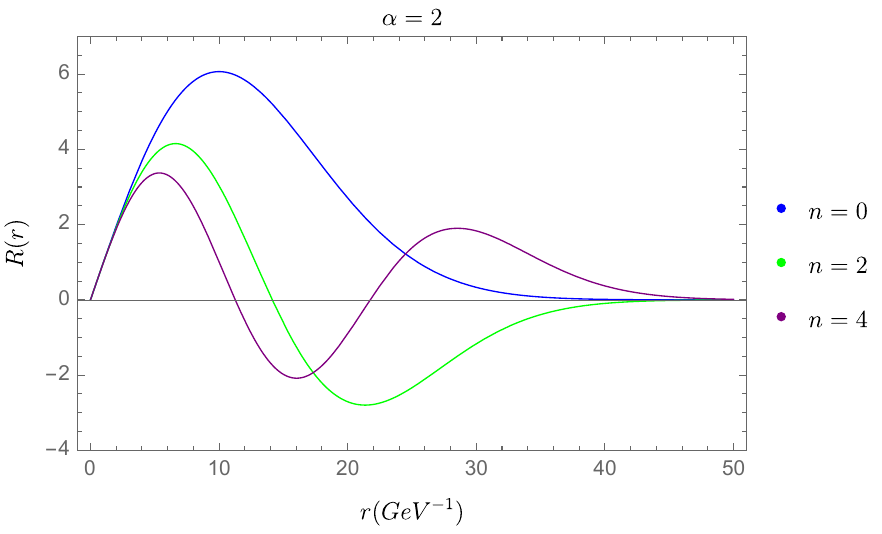}
     \end{subfigure}
        \caption{Plot of the radial function with respect to the radial coordinate $r$ in $\text{GeV}^{-1}$ for $\alpha= 1, 1.5, 2$, $\kappa=0.1$ and $\mu=0$.}
        \label{f3}
\end{figure}

\begin{figure}[H]
     \centering
     \begin{subfigure}[b]{0.3\textwidth}
         \centering
         \includegraphics[width=\textwidth]{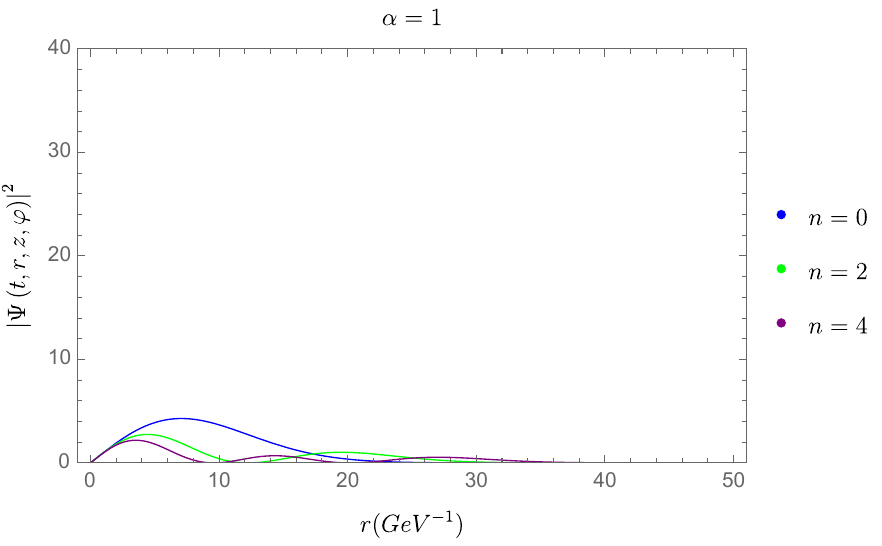}
     \end{subfigure}
     \begin{subfigure}[b]{0.3\textwidth}
         \centering
         \includegraphics[width=\textwidth]{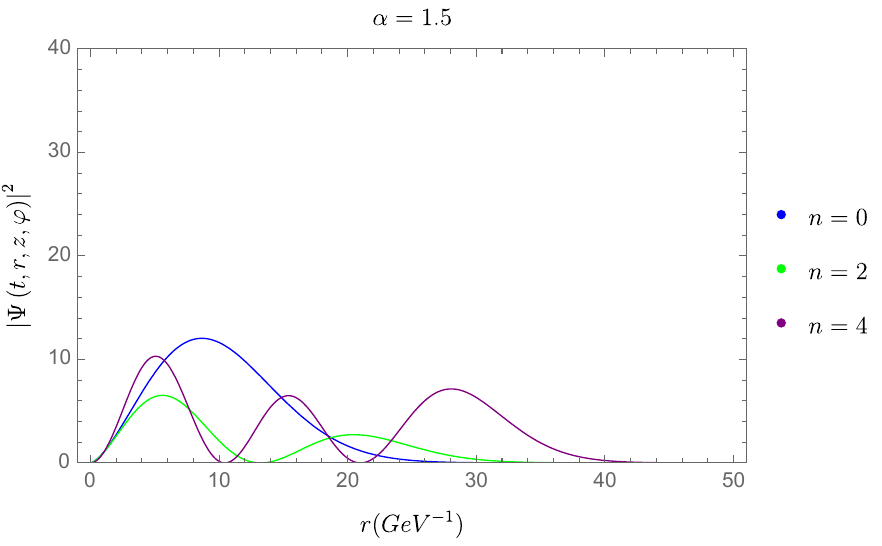}
     \end{subfigure}
    \begin{subfigure}[b]{0.3\textwidth}
         \centering
         \includegraphics[width=\textwidth]{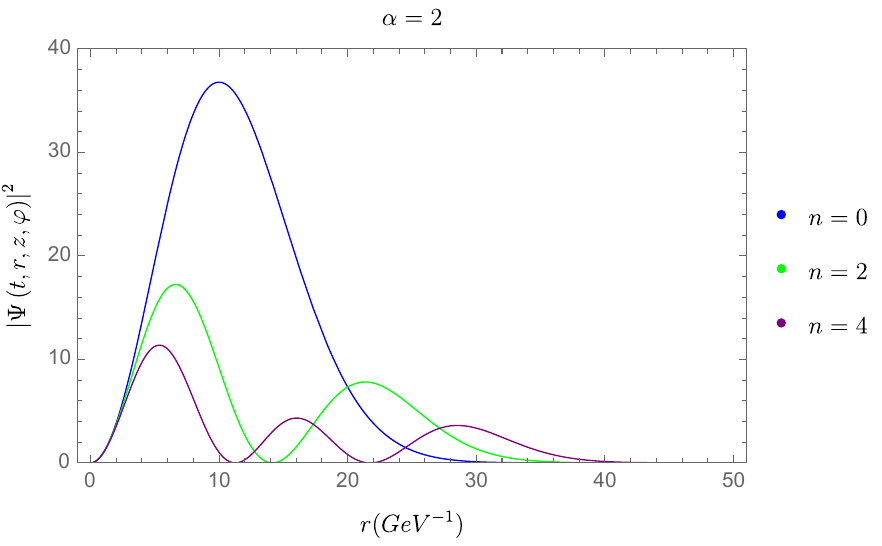}
     \end{subfigure}
        \caption{Plot of the modulus of the wave function squared with respect to the radial coordinate $r$ in $\text{GeV}^{-1}$ for $\alpha= 1, 1.5, 2$, $\kappa=0.1$ e $\mu=0$.}
        \label{f4}
\end{figure}

By applying the conditions (\ref{HeunCondI}) and (\ref{HeunCondII}) to the radial function (\ref{FRadial}), the expression for \( R(r) \) can be written as
\begin{equation}
R(r) = r^{\frac{1}{2}\alpha}e^{-\frac{1}{2}\kappa r^{2}}\text{HeunB}\left(\alpha, \beta, \alpha+2(n+1), \delta_{\mu}^{n}; \sqrt{\kappa}r\right) = r^{\frac{1}{2}\alpha}e^{-\frac{1}{2}\kappa r^{2}}P_{n,\mu}\left(\alpha, \beta; \sqrt{\kappa}r\right)
\end{equation}
where \( P_{n,\mu}\left(\alpha, \beta; x\right) \) are polynomials of degree \( n \) that satisfy the biconfluent Heun equation. Here, \( \delta_{\mu}^{n} \) denotes the Kronecker delta, and \( 0 \leq \mu \leq n \). Examples of numerical results for this expression are displayed in Fig. \ref{f3} and \ref{f4} and exhibit the microscopic aspect of the bound states, justifying the approximation considered in order to obtain eq. (\ref{eqrad2}).

\section{Conclusions}
The relevance of investigations on systems with intense magnetic fields has became much more prominent with the recent discoveries in different areas of physics, from very small systems, such as heavy ion collisions up to the cosmological level. With this motivation in mind,
in this work we studied spin-0 oscillators in a background determined by a homogeneous magnetic field and the cosmological constant $\Lambda$. In our approach, we considered a spacetime similar to the one found in \cite{vzofka2019bonnor}, \cite{barbosa2023s}, with a variation in the definition of the integration constant in order to provide an easier way to achieve an interpretation for the $\Lambda\rightarrow$ 0 limit and obtain the Minkowski spacetime.  

With the purpose of representing spin-0 bosons inside this spacetime, the Klein-Gordon equation has been written, with the inclusion of a scalar potential, solved, and the energy levels $\epsilon$ have been calculated. As an example we used a Coulomb-like potential in order to obtain a solution for the radial equation, but it is straightforward to follow the procedure presented in this work and find solutions for arbitrary forms of the potential. 
As it can be noticed in eq. (\ref{en}) these levels depend on the proposed parameters, the oscillation frequency 
$\omega$, the intensity of the magnetic field $H_0$ as well as on the potential parameter $\eta$.
Fig. \ref{f1} and \ref{f2} explore these energy levels in terms of $n$, $\ell$, $\Lambda$ and $H_0$, considering the pion as a test particle, showing how the energy spectrum is affected by these parameters. 
In Fig. \ref{f1} a high value of $n$ has been chosen, $n=1000$, and the variation of $\Lambda$ and $H_0$ on the energy levels was studied. As it can be noted, the energy increases for high values of $\Lambda$ and for some ranges of $H_0$ it is also possible to verify significant effects. Fig. \ref{f2} shows the dependence on the quantum numbers $n$ and $\ell$. There is an evident effect of $\ell$ on the energy levels showing the influence of the considered conical spacetime in the behavior of the field. The variation with $n$ is not so prominent. An interesting feature of Eq. (\ref{en}) is that, depending on the set of parameters, it may result in complex values which allows the existence of quasinormal modes for the scalar field in accordance with the results obtained in other spacetimes.
As examples, some plots for the wave functions are presented in Fig. \ref{f3} and \ref{f4}, results that indicate the existence of microscopic oscillations for the considered set of parameters.

This kind of study plays a fundamental role in many physical systems, as for example in the formulation of relativistic models to describe the structure of magnetized neutron stars, where the Landau levels are fundamental ingredients \cite{Harding_2006}, \cite{Chatterjee_2021} or even at the cosmological level.
So, we may conclude that the results presented in this work show that if we study quantum systems in arbitrary spacetimes their dynamics are determined by the structures of these spacetimes, which under specific conditions may lead to observable effects and consequently provide more insights about the relations between the general relativity and quantum mechanics.

\section{Acknowledgements}
We would like to thank CAPES (Process number: 88887.642857/2021-00) and CNPq for the financial support.

\bibliographystyle{ieeetr}
\bibliography{sample}

\end{document}